\documentclass[showpacs,twocolumn,floatfix,prb]{revtex4}
\usepackage{graphicx} 	
\usepackage{bm} 		
\usepackage{amssymb}
\usepackage{amsmath}
\usepackage{amsfonts}
\usepackage{epstopdf}

\begin{document}

\title{RKKY coupling in graphene}

\author{Annica M. Black-Schaffer}
 \affiliation{NORDITA, Roslagstullsbacken 23, SE-106 91 Stockholm, Sweden}

\date{\today}
\begin{abstract}
We study the carrier-mediated exchange interaction, the so-called RKKY coupling, between two magnetic impurity moments in graphene using exact diagonalization on the honeycomb lattice. By using the tight-binding nearest neighbor band structure of graphene we also avoid the use of a momentum cut-off which plagues perturbative results in the Dirac continuum model formulation. We extract both the short and long impurity-impurity distance behavior and show on a qualitative agreement with earlier perturbative results in the long distance limit but also report on a few new findings. In the bulk the RKKY coupling is proportional to $1/|{\bf R}|^3$ and displays $(1+\cos(2{\bf k}_D\cdot {\bf R})$-type oscillations. A-A sublattice coupling is always ferromagnetic whereas A-B subattice coupling is always antiferromagnetic and three times as large.
We also study the effect of edges in zigzag graphene nanoribbons (ZGNRs). We find that for impurities on the edge the RKKY coupling decays exponentially because of the localized zero energy edge states and we also conclude that a non-perturbative treatment is essential for these edge impurities. For impurities inside a ZGNR the bulk characteristics are quickly regained.
\end{abstract}

\pacs{75.20.Hr, 75.75.-c, 73.20.-r}

\maketitle

\section{Introduction}
Graphene is a two-dimensional honeycomb lattice of carbon and has since its isolation in 2004\cite{Novoselov04} generated a lot of attention (see e.g. Ref.~\onlinecite{CastroNeto09} and references therein).  Its two-dimensionality, linear energy dispersion, where the quasiparticles are massless Dirac fermions, and chemical potential tunable by a gate voltage are all novel features in an, essentially table-top, condensed matter system. Together with a very high mobility these properties have helped to raise the expectation of graphene being a post-silicon era candidate.\cite{Berger06, Avouris07,Geim07,Lin09} For this, functionalization of graphene using for example finite geometries, adatoms, hydrogen chemisorption, or intrinsic defects such as vacancies, has become an important goal. Adding magnetic atoms or defects have the added benefit of opening the possibility for spintronics where not only the electron charge but also its spin is actively used in devices.\cite{Wolf01, Zutic04}
One of the most important property of magnetic impurities is the effective interaction between them propagated by the conduction electrons in the bulk host, the so-called Ruderman-Kittel-Kasuya-Yoshida (RKKY) coupling.\cite{Rudermann54,Kasuya56,Yosida57} This coupling is crucial for magnetic ordering of the impurities but also offers access to the intrinsic magnetic properties of the host.

Previous work on the RKKY coupling in graphene\cite{Vozmediano05, Dugaev06, Saremi07, Brey07, Bunder09} have exclusively used a field-theory continuum model where the graphene band structure is approximated with a Dirac spectrum at each of the two inequivalent Brilloiun zone corners. Using perturbation theory the RKKY coupling has then been calculated analytically from the static spin susceptibility of this model. 
The earliest work\cite{Vozmediano05, Dugaev06} failed to recognize the importance of the bipartite lattice and predicted that the RKKY coupling is always ferromagnetic. Later it was, however, shown by Saremi\cite{Saremi07} that for all bipartite lattices at half-filling the RKKY coupling is ferromagnetic (FM) only for impurities on the same sublattice but antiferromagnetic (AFM) for impurities on different sublattices.
A perturbative approach also requires an explicit use of an ultraviolet momentum cut-off scheme for the non-interacting graphene band structure and it was recently showed that a regular sharp cut-off does not produce the correct results thus demonstrating the need for a carefully chosen regularization scheme.\cite{Saremi07} This most likely explains the discrepancies in the later results for the RKKY coupling.\cite{Saremi07,Brey07}
Furthermore, any results from a continuum model does not fully resolve the lattice structure which are likely to be important for short impurity-impurity distances $R$. In fact, the later work\cite{Saremi07,Brey07} on the RKKY coupling in graphene have only considered the long-distance limit. While this is the traditional RKKY limit, knowledge of the short distance behavior is important in nano-structures as well as when comparing with ab-initio results where the unit cell always has a finite size.

To circumvent the use of perturbation theory, the ultraviolet cut-off dependency, and to also get results for finite impurity distances we will here explicitly calculate the RKKY coupling on the honeycomb lattice using exact diagonalization in a finite system. We show that by using a large enough system it is possible to extract both short-range and long-range behavior of the RKKY coupling. We report on a qualitative agreement between our results and one of the perturbative results in the long distance limit \cite{Saremi07}, but also report on a few new findings, including a phase shift for different sublattice coupling.
This establishes not only that the standard perturbative approach to RKKY coupling in graphene is in general valid but also finally settles the issue of the exact form of the RKKY coupling.
Furthermore, we also study the RKKY coupling in zigzag graphene nanoribbons (ZGNRs) and show that the zigzag edge will significantly modify the results due to the presence of the localized zero-energy edge state.\cite{Fujita96, Nakada96, Wakabayashi99, Miyamoto99} In contrast to the bulk, a non-perturbative treatment seems to be essential for impurities along the zigzag edge. For impurities inside a ZGNR bulk-like behavior is however achieved even for narrow ribbons.

The rest of the paper is organized as follows: In the next section we introduce the model and solution method. Then we discuss the results for on-site and plaquette impurities, followed by the results for impurities on the edge and inside ZGNRs. We end with comparing our results with the previous perturbative results, and a discussion on how to experimentally achieve magnetic impurities in graphene as well as on the importance of electron-electron interactions.

\section{Method}
\begin{figure}[htb]
\includegraphics[scale = 1]{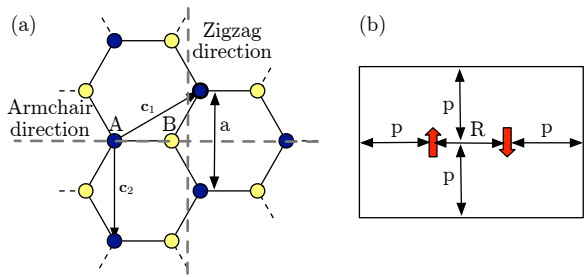}
\caption{\label{fig:graphene} (Color online) (a) The graphene honeycomb lattice with the two sublattices A and B in dark and light colors, respectively. The lattice unit vectors ${\bf c}_1$ and ${\bf c}_2$, lattice constant $a = 2.46$~\AA\, as well as the two most common directions, the zigzag and the armchair, are displayed. (b) Unit cell setup in the AFM configuration with the distance $R$ between impurity spins shown as well as the padding $p$ surrounding them.}
\end{figure}

We are here focusing on how impurity magnetic moments, or spins, interact with each other on a graphene surface. We are not concerned with the details of the interaction between the moments and graphene nor about how the moments are originally formed. We will therefore model the interaction between an impurity moment and graphene with a simple Kondo coupling term. Using the nearest neighbor  tight-binding Hamiltonian for the $p_z$-orbitals in graphene, the system can be formulated as
\begin{align}
\label{eq:H}
H =   -t \!\! \! \! \sum_{<i,j>,\sigma} \!\!\! (a_{i \sigma}^\dagger b_{j \sigma} + {\rm H.c.}) +
 J_k\sum_{i = {\rm imp}} {\bf S}_i \cdot {\bf s}_i,
\end{align}
where $a_{i \sigma}$ ($b_{i \sigma}$) annihilates an electron on sublattice A (B) in unit cell $i$ [see Fig.~\ref{fig:graphene}(a)], $<i,j>$ means nearest neighbors, and $\sigma$ is the spin index. Moreover, ${\bf S} = \pm S{\bf \hat{z}}$ is the impurity spin and  ${\bf s} = \frac{1}{2}a^\dagger_\alpha {\bf \sigma}_{\alpha \beta} a_\beta$, with ${\bf \sigma}_{\alpha \beta}$ being the Pauli matrices, is the electron spin (for sublattice B interchange $a \rightarrow b$). The constants entering are the nearest neighbor hopping in graphene $t  \sim 2.7$~eV and the Kondo coupling $J_k$ which depends on the particular impurity moment. We consider here only undoped graphene and thus no chemical potential term enters in Eq.~(\ref{eq:H}).
%

In standard RKKY perturbation theory\cite{Kittelbook} the leading interaction between two impurity moments at sites $i$ and $j$ is given by
\begin{align}
\label{eq:HRKKY}
H_{RKKY} =   J_{ij} {\bf S}_i \cdot {\bf S}_j,
\end{align}
with the effective RKKY coupling constant $J_{ij}$ proportional to the static spin susceptibility of the imbedding bulk, $J_{ij} \propto \chi^0_{ij}$.
\subsection{Exact diagonalization}
Instead of using the above perturbative result for the RKKY coupling we will calculate $J_{ij}$ by exact diagonalization of Eq.~(\ref{eq:H}) in a finite system with two impurity spins either aligned ferromagnetically (FM) or antiferromagnetically (AFM). Then $J_{ij}$ can be expressed as the energy difference between the two configurations:\cite{Deaven91} $J_{ij} = [E({\rm FM}) - E({\rm AFM})]/(2S^2)$. We will for simplicity set $S = 1$. This solution method is non-perturbative, automatically avoids any artificial ultraviolet cut-off dependencies, and is also capable of generating results for any $R$.
However, solving in a finite system creates its own problems. We apply periodic boundary conditions (PBC) to avoid the effects of edge states but then have to deal with the two impurities in one unit cell also interacting with the impurities in neighboring cells. By systemically increasing the ``padding" $p$ around the two impurities, see Fig.~\ref{fig:graphene}(b), we can determine the necessary size of the unit cell for converged results for $J_{ij}$. As seen later on in Figs.~\ref{fig:jrkky} and \ref{fig:plaq}, $p = 2R$ is in general sufficient. We have also ensured convergence with respect to the number of $k$-points in reciprocal space. 

Apart for studying the RKKY interaction in the bulk we have also looked at ZGNRs and strips. Here the graphene lattice is terminated along the zigzag direction with saturated $\sigma$-bonds whereas the $p_z$-orbital on the edge atom is unsaturated because of the one missing nearest neighbor. We have primarily studied narrow symmetric ZGNRs with width $W = 8/\sqrt{3}a$, where $a$ is the lattice constant, see Fig.~\ref{fig:graphene}.
As is well established, the zigzag graphene edge hosts localized states at zero energy which significantly changes many of the physical properties compared to the bulk.\cite{Fujita96, Nakada96, Wakabayashi99, Miyamoto99} However, the armchair edge does not have any such zero energy localized states and we find that that for large enough unit cells, the results for a ZGNR, where PBC are applied in the direction of the ribbon, are the same as those for a strip, where instead armchair edges terminate the strip.

%
\section{Results}
\begin{figure}[htb]
\includegraphics[scale = 0.85]{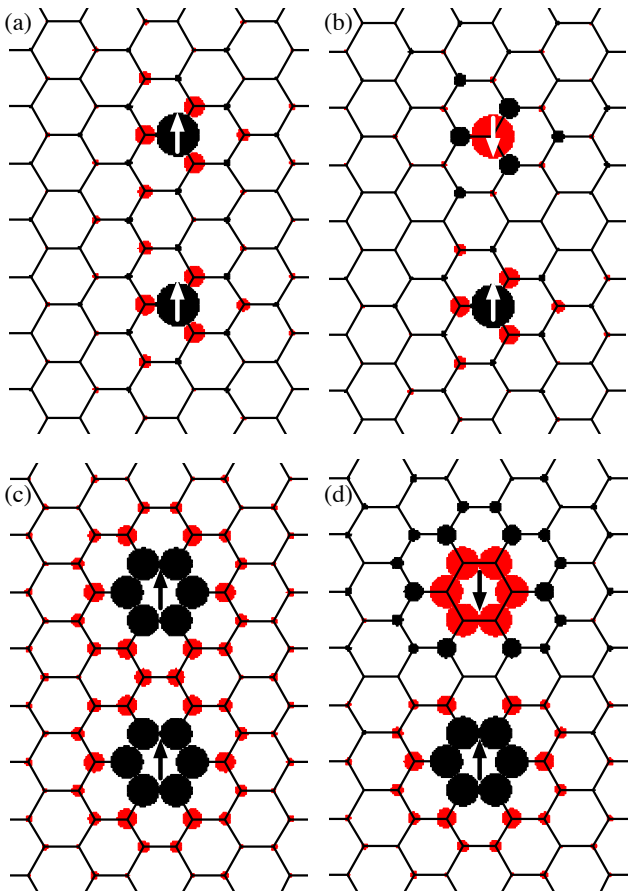}
\caption{\label{fig:spinpolbulk} (Color online) Two impurity spins along the zigzag direction positioned on-site on the B-B sublattice (a,b) and in the plaquette site (c,d) in the FM (a,c) and AFM (b,d) configurations. The impurity spins are marked with white or black arrows and the area of the circles on each site is proportional to the spin polarization, where excess spin-$\uparrow$ (spin-$\downarrow$) density is red/dark grey (black).}
\end{figure}

We start with displaying in Fig.~\ref{fig:spinpolbulk} the spin polarization, $s_A^z = (a^\dagger_\uparrow a_\uparrow - a^\dagger_\downarrow a_\downarrow)/2$ and $s_B^z$, induced into the graphene from two impurity spins positioned on-site (a,b) and in a plaquette site (c,d), in the FM (a,c) and AFM (b,d) configurations, respectively. Below we will discuss each of these results.
%
\subsection{On-site impurities}
An on-site positioning of the impurity spins, where the spins sit directly on top of an A or B atom of the graphene lattice, has  so far been the dominating setup for RKKY studies in graphene.\cite{Saremi07,Brey07,Bunder09} 
Since on-site positioning breaks the symmetry of the lattice, it is important to distinguish between A-A and A-B positioning of the two impurity spins. We have studied both of these configurations along both the zigzag and armchair directions as well as verified our predictions for the asymptotic large-$R$ behavior for several other, chiral, directions. 
Figures \ref{fig:spinpolbulk}(a,b) show typical spin polarization patterns of two impurity spins both on the B sublattice and along the zigzag direction. We clearly see that the spin polarization has different signs on the two sublattices close to an impurity and therefore one would expect A-A (B-B) impurities to prefer a FM coupling whereas AFM coupling should be the case for A-B (B-A) impurities.

\begin{figure}[htb]
\includegraphics[scale = 0.95]{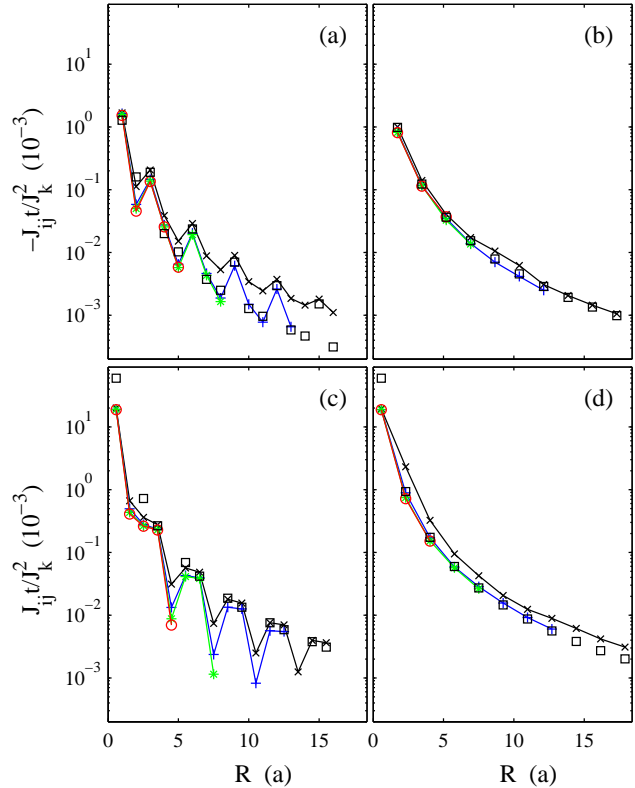}
\caption{\label{fig:jrkky} (Color online) $-t/J_k^2 J_{ij}$ for A-A (a,b) and  $t/J_k^2 J_{ij}$ for A-B (c,d) on-site positioning of the impurity spins along the zigzag (a,c) and armchair (b,d) directions as function of impurity distance $R$. Solid curves represent calculated results with padding $p = 1R-4R$ (black $\times$, blue $+$, green $\star$, red $\circ$), where the lines are only a guide to the eye, whereas the large-$R$ dependence in Eqs.~(\ref{eq:JAA}-\ref{eq:JAB}) with $C = 1/(72\sqrt{3}\pi)$ is displayed with black $\square$s.
}
\end{figure}

Figure \ref{fig:jrkky} shows $J_{ij}$ as a function of the impurity distance $R$ for all four different configurations of sublattice and zigzag/armchair directions for on-site impurities. 
%
First of all we can directly verify that the RKKY coupling is always FM for A-A sites and AFM for A-B sites as seen in the sign difference of $J_{ij}$. This has been predicted before \cite{Saremi07,Brey07} and is a property of the bipartite lattice.\cite{Saremi07}
Secondly, we conclude that padding $p = 2R$ is enough for well converged results. 
Third, we extract the following functional dependence in the large $R = |{\bf R}|$ limit: 
\begin{align}
\label{eq:JAA}
J_{ij,{\rm A-A}}({\bf R}) & = -C\frac{J_k^2}{t}\frac{1 + \cos(2{\bf k}_D\cdot {\bf R})}{|{\bf R}|^3} \\
\label{eq:JAB}
J_{ij,{\rm A-B}}({\bf R}) & = C\frac{3J_k^2}{t}\frac{1 + \cos(2{\bf k}_D\cdot {\bf R} + \pi)}{|{\bf R}|^3},
\end{align}
for $R$ measured in units of the lattice constant $a$. In Fig.~\ref{fig:jrkky} these results are plotted as black $\square$s using the numerical prefactor $C = 1/(72\sqrt{3}\pi)$ and, as seen, there is essentially a perfect agreement at larger $R$.
In the above equations ${\bf k}_D$ is the reciprocal vector for the Dirac points, i.e. the corners of the Brillouin zone. There are six such vectors and for the A-A configuration the result is independent on the particular choice of ${\bf k}_D$ since ${\bf R}$ is then a lattice translation vector, i.e.~${\bf R} = n_1 {\bf c}_1 + n_2 {\bf c}_2$ where $n_1,n_2$ are integers. However, for the A-B configuration ${\bf R}$ is not a lattice translation vector and $\cos(2{\bf k}_D\cdot {\bf R})$ can then depend on the choice of ${\bf k}_D$. Eq.~(\ref{eq:JAB}) is only correct when ${\bf k}_D\cdot {\bf R}$ is maximized, i.e. when ${\bf k}_D$ is chosen to be as parallel as possible to ${\bf R}$.
Eqs.~(\ref{eq:JAA}-\ref{eq:JAB}) are very similar to the results derived earlier by Saremi \cite{Saremi07} using perturbation theory in a continuum model except for the addition of the $\pi$-phase factor and the need for specifying ${\bf k}_D$ in the A-B result, as just discussed. The $\pi$-phase factor is essential for reproducing the numerical results for A-B sublattice coupling as these are $180^\circ$ out of phase with the A-A results. These two discrepancies stem from the fact that in all previous work ${\bf R}$ has improperly been treated as a lattice translation vector even for A-B impurities. As seen in our results, choosing ${\bf R}$ to be the correct impurity-impurity distance not only changes the RKKY coupling at short distances, as one might have expected, but also produces the $\pi$-phase shift and a need for an explicit choice of ${\bf k}_D$ for all ${\bf R}$. We strongly believe that a proper handling of ${\bf R}$ in a perturbative treatment in the continuum model will also include these two additional corrections found in our numerical results.
Our numerical prefactor $C = 1/(72\sqrt{3}\pi)$ also agrees with the results of Saremi\cite{Saremi07} if one explicitly use a factor of $t = 8/3 \approx 2.67$~eV to produce the proper energy dimension of their results.
We thus conclude that our non-perturbative results agree, up to a few small, and traceable differences, with the perturbative results of Saremi and they firmly establish that the RKKY coupling is oscillatory for certain ${\bf R}$-vectors, although never changes sign on the same sublattice, which is contrary to some other recent RKKY results.\cite{Brey07}
However, note that due to the impurities only appearing at lattice sites, the $(1 + \cos(2{\bf k}_D\cdot {\bf R}))$-oscillation is in general undersampled. For the zigzag direction the period is $3a$ instead of $3a/4$ whereas for the armchair direction the period is infinitely long. Also, for the cases reported in Fig.~\ref{fig:jrkky}, it is only for the zigzag A-B configuration that the RKKY coupling completely disappears at certain sites.
What makes the RKKY coupling in graphene unusual is this non-sign changing oscillation on the same sublattice as well as the $1/R^3$ decay as compared to $J_{ij} \propto \sin(2k_F R)/(2 k_F R)^2$ for an ordinary 2D metal.\cite{Fischer75,Beal87}
Finally, we are not only able to extract the long-distance behavior but can also directly see the deviations from this behavior at short impurity distances. As seen in Fig.~\ref{fig:jrkky}, the results along the zigzag direction are well converged toward the large-$R$ results around $R\sim 10a$. The exceptions are the results for every third lattice site in the A-B configuration which are zero in Eq.~\ref{eq:JAB}.
 Along the armchair direction the convergence is even faster and, except for nearest neighbor impurity spins, Eqs.~(\ref{eq:JAA}-\ref{eq:JAB}) give correct results for any $R$. We thus conclude that the large-$R$ limit is reached for surprisingly small impurity distances $R$.

\subsection{Plaquette impurities}
\begin{figure}[htb]
\includegraphics[scale = 0.95]{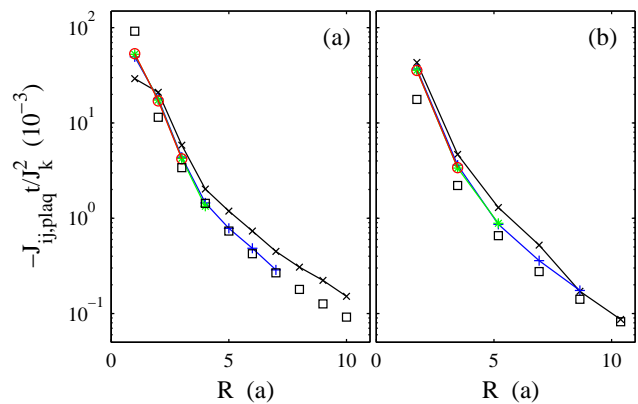}
\caption{\label{fig:plaq} (Color online) $-t/J_k^2 J_{ij,{\rm plaq}}$ for impurities in the plaquette site along the zigzag (a) and the armchair (b) direction. Same color coding as in Fig.~\ref{fig:jrkky}.}
\end{figure}

For magnetic atoms deposited on graphene, the on-site position might not the most energetically favorable but instead the atoms can prefer to sit in the middle of the hexagon, in the plaquette site.\cite{Mao08,Wehling09} We will here first study the simple situation where the impurity spin couples incoherently, and with the same coupling constant $J_k$, to all six nearest neighbors in the honeycomb lattice. Representative spin polarization maps in this case are shown in Figs.~\ref{fig:spinpolbulk}(c,d) which have plaquette impurities along the zigzag direction.
For this incoherent, symmetric, coupling to the lattice the large-$R$ result can in fact be directly derived from the on-site results in Eqs.~(\ref{eq:JAA}-\ref{eq:JAB}) by summing the interactions between all combinations of nearest neighbors of each impurity spin. Doing so, it turns out that in this case the oscillations cancel, the AFM coupling prevails, and the asymptotic large-$R$ result is given by
\begin{align}
\label{eq:Jplaq}
J_{ij,{\rm plaq}}({\bf R})  = C\frac{36J_k^2}{t}\frac{1}{|{\bf R}|^3}.
\end{align}
This is the same result as derived by Saremi \cite{Saremi07} despite the additional $\pi$-phase factor in Eq.~(\ref{eq:JAB}). 
Figure \ref{fig:plaq} shows our exact diagonalization results together with Eq.~(\ref{eq:Jplaq}). As before, we see that $p = 2R$ is enough to reach converged numerical results. For plaquette impurities along the armchair direction, Fig.~\ref{fig:plaq}(b), there is a systematic increase in $J_{ij,{\rm plaq}}$ for small $R$ compared to the long-distance result but the asymptotic behavior is nonetheless reached before $R = 10a$. For the zigzag direction, Fig.~\ref{fig:plaq}(a), there is some oscillations around the asymptotic value for short $R$ but convergence with respect to this value is reached already at $R \sim 4a$.
%

While the incoherent case above is straightforward to derive from the results of on-site impurities, coherent coupling, where of a plaquette impurity spin couples to a linear coherent combination of the nearest neighbor conduction electrons, is physically more realistic. Saremi\cite{Saremi07} showed that the same cancelation of the oscillations that occur in Eq.~(\ref{eq:Jplaq}) also makes the $1/R^3$ contribution disappear altogether for coherent coupling. Since we get the same cancellation in the incoherent case when we properly treat the impurity distance, we conclude that this conclusion is still valid. Coherent plaquette impurities thus have a significantly weaker, and thus less relevant, RKKY coupling than the other configurations studied here.
\subsection{Impurities in ZGNRs}
Zigzag edges in graphene have proven to be an exciting new playground for magnetism since this termination leads to a localized, flat-band, zero energy edge state and is thus extremely prone to spin polarization.\cite{Fujita96, Nakada96, Wakabayashi99, Miyamoto99} This zero energy singularity in the local density of states (LDOS) has been claimed to significantly change the perturbative RKKY coupling result because of a qualitatively modified behavior of the spin susceptibility at the edge.\cite{Bunder09}  It is therefore of large interest to further study these systems and establish the consequences of zigzag edges on the RKKY behavior in an exact, non-perturbative setting. We have found that an armchair edge does not significantly effect the RKKY coupling and thus it is only necessary to study the zigzag edge in order to establish the qualitative effect of edges on the RKKY coupling. 

\begin{figure}[htb]
\includegraphics[scale = 0.72]{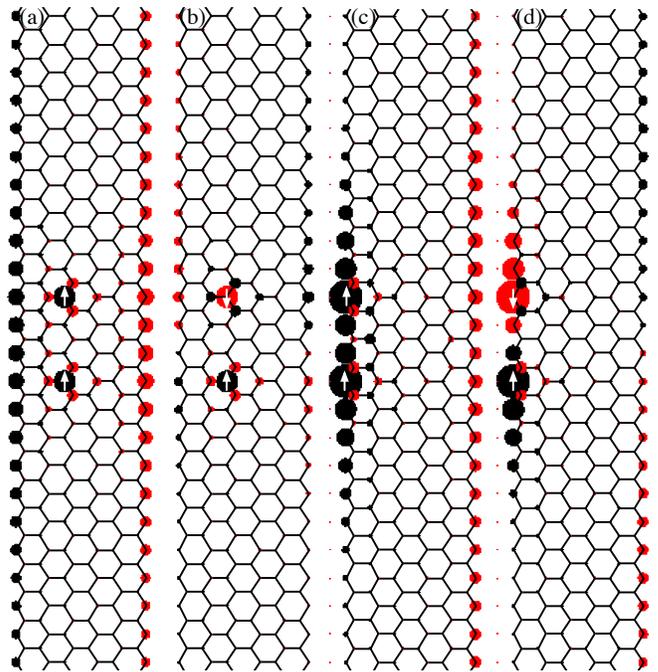}
\caption{\label{fig:spinpolnr} (Color online) Two on-site impurity spins inside a narrow ZGNR (a,b) and on the edge of the same ZGNR (c,d) in the FM (a,c) and AFM (b,d) configurations, respectively. Same color coding as in Fig.~\ref{fig:spinpolbulk}.}
\end{figure}

\begin{figure}[htb]
\includegraphics[scale = 0.95]{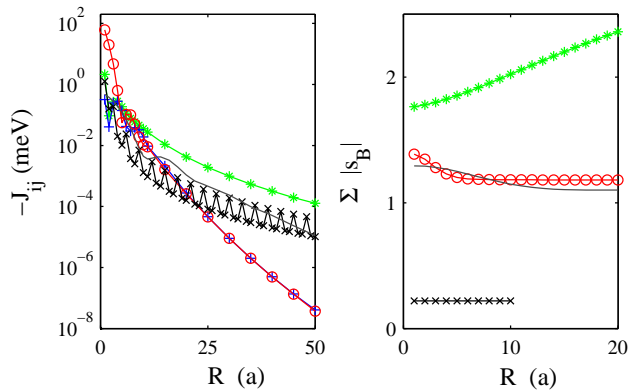}
\caption{\label{fig:edge} (Color online) (a) $-J_{ij}$ for impurities along the zigzag direction on the zigzag edge of a narrow ZGNR (red, $\circ$), inside the same ZGNR (green, $\star$), and the asymptotic behavior in the bulk (black, $\times$) as function of impurity distance $R$. In (blue, $+$) is $J_{ij}$ for impurities on opposite edges of the ZGNR. Here $t = J_{k} = 1$ expect the gray line where $t = 1, J_k = t/100$.
(b) The total spin polarization in sublattice B, $\sum |s_B|$, for the same situations as in (a) in the FM impurity configuration. A large enough unit cell was used to ensure convergence of $\sum |s_B|$ in the $R$-range displayed.
}
\end{figure}

Figure \ref{fig:spinpolnr} shows the spin polarization for two prototypical situations in a narrow ZGNR: impurities inside the ZGNR (a,b) and along the edge (c,d).
%
%
The RKKY coupling $-J_{ij}$ for the configurations in Fig.~\ref{fig:spinpolnr} is shown in Fig.~\ref{fig:edge}(a) for the parameters $t = J_k = 1$. While this is a rather unphysically high value for $J_k$ it helps displaying the essential features in a numerically accessible $R$-range. We have for comparison also included the results for $J_k = t/100$, and, as discussed below, the same physical behavior is governing both $J_k$-values, although all significant features get extended over a longer $R$-range for smaller $J_k$, making a complete numerical study harder.
%
For reference in Fig.~\ref{fig:edge}, the asymptotic zigzag result in the bulk is shown in black and we directly see that impurities in narrow ZGNRs behave significantly different. The coupling between edge impurities (red, $\circ$) is larger than in the bulk for small distances and also displays some oscillations in that regime, but then decays exponentially for larger $R$ which is in sharp contrast to the power-law decay in the bulk.
Displayed is also $J_{ij}$ for the RKKY coupling between opposite edges, the opposite sign being a consequence of the two edges belonging to different sublattices. As seen, after an initial short distance, where the width of the ZGNR is important, same edge and opposite edge impurities couple with equal strength. This is another difference compared to the bulk results where the AFM coupling is 3 times larger than the FM coupling. We thus draw the conclusion that the edges are dominating the response for edge impurities. This could have been anticipated already from Figs.~\ref{fig:spinpolnr}(c,d) where the spin polarization is seen to be large only in the vicinity of the impurity and only along the edge, it does not spread significantly into the bulk. 
The essential features of the RKKY coupling along the edge can be understood from the following argument: A magnetic impurity will always force a spin polarization of the graphene in its vicinity. For a zigzag edge impurity this spin polarization can trivially, and essentially without energy penalty, be achieved by polarizing the localized edge state at zero energy without much polarization of the surrounding bulk. Consequently, away from the impurity, the edge state will also quickly become unpolarized, much more quickly than the bulk can lose its spin polarization, and thus the RKKY coupling between two edge impurity spins decays much faster than in the bulk. With this argument one would expect the total amount of polarization in the B sublattice for the FM configuration, $\sum |s_B|$, to be large for small $R$, as then the two impurities interact very strongly with each other, but then  rapidly decay with $R$ until its flattens out to a value equal to the sum of the total spin polarization of two uncoupled edge impurity spins. This behavior is confirmed in Fig.~\ref{fig:edge}(b) where the asymptotic value is shown to be reached already around $R =7a$ for $J_k = t$. This should be compared to the evolution of the spin polarization for the equivalent configuration in the bulk where the total spin polarization is almost constant, as displayed by the black curve in Fig.~\ref{fig:edge}(b).
The exponential decay of $J_{ij}$ is in sharp contradiction to earlier calculations on the same width ZGNR by Bunder {\it et al.}\cite{Bunder09} who used an analytical approach to the tight-binding structure of graphene to calculate the spin susceptibility and thus obtained perturbative results for the RKKY coupling. Both methods predict an equivalence between same and opposite edge impurity spins but Bunder {\it et al.} report an almost linear decay with distance for small distances followed by sign oscillations in the RKKY coupling. 
Despite a thorough investigation of the RKKY coupling for $R \leq 130a$, we were not able to detect any deviations from the exponential decay, and thus no sign changes, within the numerical accuracy which was a factor of $10^{-11}$ of the RKKY coupling at $R \sim 0$. We thus conclude that when studying these edge states, a non-perturbative method appears to be essential when calculating the RKKY coupling.

In the bulk $J_{ij} \propto J_k^2$ and $\sum |s_B| \propto J_k$ but for edge impurities these simple relations do not longer hold. For all $J_k \geq t/10$ we have found that the asymptotic large-$R$ induced spin polarization from edge impurities is $\sum |s_B| \approx 1.1$ and even for $J_k =t$ the asymptotic value of $\sum |s_B|$ is only slightly higher due to a finite amount of induced polarization in the nearby bulk, see Fig.~\ref{fig:edge}(b). Thus even in the limit of vanishing $J_k$, edge impurities are going to elicit a finite spin polarization response of the ZGNR. This is again due to the extreme easiness of polarizing the zero energy edge state. However, when $J_k$ decreases the spin polarization per edge site naturally goes down so the polarization now instead have to be spread over more edge sites. This has an interesting consequence for the RKKY coupling: When $J_k$ decreases, the more elongated polarization response causes the oscillations in the RKKY coupling for small $R$ to be spread out over a longer $R$-range. Eventually, however, an exponential decay is achieved for large enough $R$ for any $J_k$, as also seen in the gray curve in Fig~\ref{fig:edge}(a) for $J_k = t/100$. But note that the exponential decay rate also becomes smaller when the spin polarization gets more elongated along the edge, thus making the RKKY coupling at large $R$ {\it larger} for decreasing $J_k$. As a direct consequence, the RKKY coupling between edge impurities is going to be larger than the coupling between bulk impurities over a larger $R$-distance the smaller the $J_k$. Of course in the extreme large-$R$ limit the exponential decay is always going to make the edge impurity RKKY coupling smaller than the bulk coupling. 
It is worth pointing out before leaving the treatment of same edge impurities that for all of the effects described above, the width of the ribbon is not essential and we thus expect the same behavior even for much wider ribbons.

Physically in between zigzag edge and bulk impurities we find impurities positioned inside a narrow ZGNR. The spin polarization for this situation is displayed in Fig.~\ref{fig:spinpolnr}(a,b).
Here the bulk is polarized in the direction parallel to the ZGNR but this bulk polarization also induces an edge polarization. Thus the absolute value of the polarization is here very large and it grows with the distance $R$ since all the bulk between the two impurities is at least lightly polarized and any amount of bulk polarization is going to elicit a large polarization of the edge state. This behavior is clearly seen in Fig.~\ref{fig:edge}(b) where the total spin polarization increases linearly with $R$.
However, the edge states are also very easily unpolarized, as discussed above, so the effect on the RKKY coupling from the edges is not expected to be very large compared to the bulk contribution. This prediction is confirmed in Fig.~\ref{fig:edge}(a) where we see that, while the coupling is somewhat larger than in the bulk and lacks oscillations, it decays with approximately the same power-law as in the bulk. Also, $J_{ij} \propto J_k^2$ and $\sum |s_B| \propto J_k$ for these impurities, which further establish the bulk-like behavior of impurities inside a narrow ZGNR. We anticipate that for wider ZGNRs all bulk properties are fully recovered. 

%
\section{Discussion}
Our results above are obtained without any approximations except the finite size of the system, assuming a non-interacting picture for the electrons in graphene, and using a nearest-neighbor hopping band structure.
The first approximation we have taken care to handle systematically and as long as the padding $p$ around the impurities are twice or larger than the impurity-impurity distance $R$ in each direction, the results are well converged. Also, since the asymptotic behavior is reached for relatively small $R$ we have been able to extract the large $R$ limit to make a comparison with earlier, partly disagreeing, results\cite{Vozmediano05, Dugaev06, Saremi07, Brey07} obtained using a perturbative approach within the continuum field-theory model. Eqs.~(\ref{eq:JAA})-(\ref{eq:JAB}) are our results for large $R$ for bulk impurities and these are closely related to one of the results in the literature\cite{Saremi07} although a $\pi$-phase factor and the need for explicitly choosing $k_D$ for the A-B configuration are new findings. This difference is most likely due to a previous improper treatment of the impurity distance for impurities on different sublattices. Thus our results establish both that the usual perturbative treatment of RKKY coupling is appropriate in the bulk and, that, while the RKKY coupling in graphene does not undergo sign changes with distance on the same sublattice, it still has a $(1+\cos(2{\bf k}_D \cdot {\bf R}))$-oscillation, a fact that has only been pointed out in one of the previous works.\cite{Saremi07}
Despite these oscillations the facts that the AFM A-B coupling is 3 times larger than the FM A-A coupling and that the coupling is stronger at short distances result in a strong tendency towards AFM order for any random configuration of impurities. Thus these new results still support the conclusion of magnetic defects in graphene creating a dilute antiferromagnet at low enough temperatures.\cite{Brey07} The moments will here be oriented in opposite directions on the two sublattices with the total magnetic moment equal to zero. Since the RKKY coupling is always AFM (incoherent coupling) or very small (coherent coupling) for plaquette impurities a dilute AFM state could also be present for plaquette impurities.
We have also studied ZGNRs where we have shown that the zero energy edge state present on the zigzag graphene edge significantly modifies the coupling between impurity spins. We show that defects along the edges couple very strongly at short distances but that the coupling finally decays exponentially with distance in contrast to the $R^{-3}$ power-law decay in the bulk. The exponential decay is a  consequence of the extreme easiness of polarizing and unpolarizing the localized zero energy state. This result disagrees with earlier perturbative results\cite{Bunder09} where they found the decay to be almost linear for small $R$ followed by oscillatory sign changes. 
So while the standard perturbative RKKY treatment is approximately valid in the bulk we have here shown that for ZGNR edge impurities a numerical non-perturbative treatment is essential.  

The approximation of ignoring electron-electron interaction in graphene is on the other hand harder to motivate. The non-interacting picture has been predominant in the study of RKKY coupling\cite{Dugaev06, Saremi07, Brey07, Bunder09} as well as in other theoretical work on magnetic adatoms in graphene.\cite{Uchoa08, Cornaglia09, Uchoa09, Zhu09} However, there seem to be growing evidence for the importance of electron-electron interactions in graphene with theoretical results pointing to graphene being close to a Mott insulator state.\cite{Drut09,Drut09b,Drut09c, Khveshchenko01b, Khveshchenko04, Gorbar02, Herbut06, Herbut09, Herbut09b} Electron interactions could thus potentially significantly modify the magnetic properties of graphene and therefore also the RKKY coupling. One such example would be if the insulating state is a N\'eel phase.\cite{Herbut06} 
While a comprehensive treatment of electron interactions in graphene is extremely hard, a density functional theory (DFT) calculation of a carefully chosen system should be able to offer insight into the importance of interactions for the RKKY coupling. However, not only will the magnetic moment and its coupling to graphene be more complicated in a real system than in the idealized Eq.~(\ref{eq:H}) but also a large unit cell is needed, making the DFT calculation highly computationally intensive.
Nonetheless, some DFT results exist for short distances\cite{Santos09} and line defects,\cite{Pisani08} both pointing to a slower power-law decay than $R^{-3}$. A detailed comparison of such results with our short distance RKKY coupling may offer insights in the applicability of the non-interacting approximation and is the goal of a future study. In fact, initial data points to the possibility of electron-electron interactions producing strong enough corrections to the non-interacting results that it is reasonable to then ignore any corrections to the nearest neighbor hopping band structure in comparison.
In a ZGNR the effect of electron-electron interaction is possibly larger than in the bulk as here any electron-electron interaction will drive a spontaneous spin polarization of the edges.\cite{Hikihara03, Lee05, Pisani07, Rossier08, Yazyev08b, Feldner09} Thus, in a real ZGNR the edge is already polarized and an impurity spin can then not as easily polarize the edge as found in the non-interacting picture, especially since $J_k$ is usually a small parameter. 
A recent DFT study\cite{Soriano09} has shown on an almost exponentially decaying RKKY coupling for nearby impurities inside a narrow ZGNR which points to the importance of electron-electron interactions even inside ZGNRs. This would be in contrast to the non-interacting picture where impurities inside a ZGNR behave essentially bulk-like. Detailed results on the influence of electron-electron interactions for ribbons is also a subject of the future study.

We have here also explicitly ignored the issue of how an external magnetic moment is created in graphene, but for any experimental verification or use of our results such considerations have to be taken into account. The simplest implementation is probably to deposit a magnetic adatom on top of graphene, such as Co or Fe.\cite{Mao08,Wehling09} We also expect the long distance behavior to be qualitatively similar for substitutional magnetic atoms, of which at least Co has been shown to be magnetic in graphene.\cite{Santos09} Single-atom vacancies and hydrogen chemisorption defects have also both been shown to possess a magnetic moment in graphene\cite{Lehtinen04, Yazyev07, Yazyev08, Pisani08, Palacios08, Kumazaki08} and should behave similarly to substitutional magnetic atoms. 
In fact, for vacancies, Lieb's theorem\cite{Lieb89} gives the same AFM/FM state as found here. It has even been proposed that the FM state found in proton-radiated graphite\cite{Esquinazi03, Ohldag07} is due to excess vacancy creation on one sublattice, which is in agreement with our findings.\cite{Yazyev08}
At very short impurity distances, however, annihilation of vacancies as well as direct exchange between magnetic moments can also be of importance. 
In aggregate, there exist multiple possibilities for experimentally studying the RKKY coupling in graphene. 
Moreover, the critical coupling for the Kondo effect is likely too high in undoped graphene\cite{Sengupta08, Cornaglia09, Zhu09, Wehling09} and therefore the Kondo effect should not compete with the RKKY coupling in any of these systems.

In summary we have studied the RKKY coupling between two impurity spins on graphene for any impurity distance $R$ using exact diagonalization. Our results largely agree with an earlier perturbative result in the large $R$ limit where $J_{ij} \propto \pm(1 + \cos(2{\bf k}_D\cdot {\bf R}))/|{\bf R}|^3$, with A-A sublattice arrangement FM and A-B sublattice arrangement AFM, although an additional $\pi$-phase shift for the A-B sublattice arrangement is a new finding.
The large $R$ limit is reached within a few lattice unit constants and the deviations are in general not large even for small $R$ in the bulk. We have also studied the effect of zigzag edges and found that impurities along this edge display an exponentially decaying coupling at large $R$ due to the easiness of polarization of the zero energy edge, state in an non-interacting picture. This result for edge impurities is however in contrast with earlier perturbative results,\cite{Bunder09} pointing to the importance of an exact, non-perturbative, treatment of the RKKY coupling in ZGNRs. Impurities away from the edge, even in narrow ZGNRs, regain most of the bulk characteristics.

%
\begin{acknowledgments}
The author thanks Sebastian Doniach, Jonas Fransson, and Biplab Sanyal for valuable discussions and Eddy Ardonne for helpful comments on the manuscript.
\end{acknowledgments}


\end{document}